\documentclass[aps,prb,twocolumn,floatfix,amsmath,amssymb]{revtex4}
\usepackage[final]{graphicx}
\usepackage{bm}
\usepackage{float}
\usepackage{afterpage}

\newcommand\vex[1]{\boldsymbol{#1}}

\newcommand{\cH}{{\cal H}}

\newcommand{\eV}{\, \mbox{eV}}

\newcommand{\bite}{$\mbox{Bi}_{2}\mbox{Te}_{3}$\,}
\newcommand{\bise}{$\mbox{Bi}_{2}\mbox{Se}_{3}$\,}
\newcommand{\biseFe}{$\mbox{Bi}_{2-x}\mbox{Fe}_{x}\mbox{Se}_{3}$\,}

\newcommand{\be}{\begin{eqnarray}}
\newcommand{\ee}{\end{eqnarray}} 

\begin{document}
\bibliographystyle{plain}

\title{Surface magnetic ordering in topological insulators with bulk magnetic dopants}
\author{G. Rosenberg}
\affiliation{Department of Physics and Astronomy, University of British Columbia, Vancouver, BC, Canada V6T 1Z1}
\author{M. Franz}
\affiliation{Department of Physics and Astronomy, University of British 
Columbia, Vancouver, BC, Canada V6T 1Z1}
\affiliation {Kavli Institute for Theoretical Physics, University of California, Santa Barbara, CA 93106-4030, USA}

\begin{abstract}
We show that a three dimensional topological insulator doped with magnetic
impurities in the bulk can have a regime where the surface is magnetically ordered
but the bulk is not. This is in contrast to conventional materials where bulk ordered
phases are typically more robust than surface ordered phases. The
difference originates from the topologically protected gapless surface states
characteristic of topological insulators. We study the problem using a
mean field approach in two concrete models that give the same
qualitative result, with some interesting differences. Our findings could
help explain recent experimental results showing the emergence of
a spectral gap in the surface state of Bi$_2$Se$_3$ doped with Mn
or Fe atoms, but with no measurable bulk magnetism.
\end{abstract}

\maketitle

\section{Introduction}

In recent years, the field of topological insulators (TI) has attracted much attention and research in the condensed matter community.\cite{ReviewHasanKane,ReviewMoore,ReviewQiZhang} The advance has been rapid, on both the theoretical and experimental fronts, however, many challenges still remain. Among these perhaps the most important is gaining experimental control over the bulk and surface conduction in three-dimensional TIs. In this paper we address one aspect of this challenge that is associated with a magnetically induced excitation gap in the topologically protected surface states.

\begin{figure}
\includegraphics[width=2.5in]{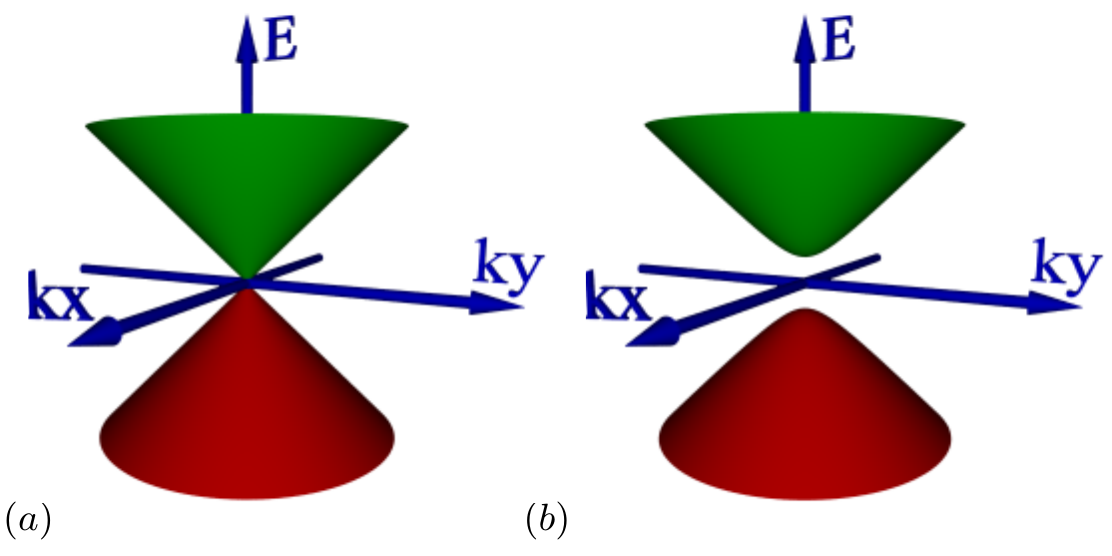}
\caption{(a) Energy spectrum of a massless Dirac fermion. The bottom (red) cone is the valence band, fully occupied at half filling. The top (green) cone is the conduction band, which is assumed to be vacant. (b) Massive Dirac fermion. This figure shows how the opening of a gap tends to lower the free energy.}
\label{FigDiracCone}
\end{figure}

Topological insulators  are bulk insulators in 2D or 3D with strong spin orbit coupling (SOC) and protected gapless surface states.\cite{KaneMele1,KaneMele2, BernevigZhang,KonigHgTe,ZhangHgTe,FuKane} The topological protection of the surface states arises due to time reversal invariance (TRI). The surface states are conducting, have a characteristic linear (Dirac) dispersion and exhibit spin-momentum locking. 
The most studied and most promising 3D TI is the semiconducting thermoelectric \bise, with a relatively large band gap of $\sim 0.3$eV and a simple surface state consisting of a single Dirac cone.\cite{ZhangSingleDiracCone, XiaLargeGap} The spectrum of \bise and other TIs has been studied using angle resolved spectroscopy\cite{hsieh1,chen1,xia1} (ARPES) and scanning tunnelling microscopy,\cite{hanaguri1,roushan1,zhang1} showing that the surface states form an almost ideal Dirac cone, illustrated in Fig.~\ref{FigDiracCone}a, familiar from studies of graphene.\cite{Neto}

Breaking TRI, for example by adding magnetic dopants (as we shall discuss), is expected to open a gap in the spectrum of the surface states. The resulting spectrum then resembles that of a `massive' Dirac fermion (Fig.~\ref{FigDiracCone}b). There is considerable interest in having a system with an odd number of massive Dirac fermions, since it is predicted to exhibit many interesting topological phenomena, including the half quantum Hall effect on the surface ($e^{2}/2h$ Hall conductance),\cite{EssinMagneto} the image magnetic monopole (an electric charge adjacent to a TI results in the field of a magnetic monopole embedded in the TI),\cite{QiMonopole,ZangMonopole} and a Kerr/Faraday angle quantization in units of the fine structure constant.\cite{tse1,maciejko1} A tunable gap would also allow the control of the surface transport, and could in addition lead to unique practical applications associated with purely electric control of the surface magnetization.\cite{nagaosa1,garate2}

A signature of the massive Dirac fermion has been observed recently using ARPES in magnetically doped \bise,\cite{ChenMassiveDirac,WrayCoulomb} although the interesting effects associated with it have yet to be seen in a laboratory.
A surprising feature of these experiments is that the gap in the surface spectrum appears {\em without} bulk magnetic ordering, even though the dopants are uniformly distributed everywhere in the 3D sample. These findings raise several important questions concerning the precise conditions under which TRI-breaking perturbations open up a gap.  Can a gap open in the surface state of a TI in a TRI-broken phase which however lacks global magnetic ordering? Although we know of no systematic study of this problem, simple arguments suggest that unordered magnetic moments do not open a gap. Consider creating such a disordered state from a uniform 2D ferromagnet (FM) in the surface of a TI by introducing domains with opposite magnetization (taken to point in the direction perpendicular to the surface). It is well known that the resulting domain walls carry topologically protected gapless fermionic modes.\cite{FuMaj} As the number of the domains grows so does the density of the low-energy fermion modes, ultimately presumably recovering the 2D gapless state characteristic of the system with unbroken TRI. The above argument thus suggests that uniform magnetic ordering over large domains is necessary to gap out the surface modes in a TI.

In this paper, we lay out the hypothesis that a temperature window exists in which the surface of a magnetically doped TI is magnetically ordered but the bulk is not. We present a simple and intuitive argument why this is so, and back it up by a mean-field calculation for two simple tight binding TI models: a cubic-lattice regularized \bise and a model on the perovskite lattice. Our results show that indeed a sizeable regime such as described above could exist in real TIs, and this indicates a possible physical explanation for the results seen in experiments.\cite{ChenMassiveDirac,WrayCoulomb}


\section{Surface Magnetic ordering in topological insulators} 
\subsection{Surface doping}

The most natural way to attempt to open up a gap in the surface state of a TI is to coat the surface with a ferromagnetic material, with magnetization perpendicular to the surface. Theoretically, this causes a gap to open up, proportional to the magnetization of the FM coating.\cite{FuMaj,QiTheoryTRI} To illustrate this point, consider the effective low energy Hamiltonian for electrons on the surface of a 3D TI that lies parallel to the $x$-$y$ plane
\be
 \cH_{0} = v (k_{x}\sigma_{y} - k_{y}\sigma_{x})
 \label{HSurface}
\ee
where $v$ is the Fermi velocity, and $\sigma_{i}$ are Pauli matrices in the spin subspace. If we coat this surface with a ferromagnetic coating with magnetization $\vex{M}=M\hat{z}$ then we get an additional term in the Hamiltonian
\be
\cH = \cH_{0} + JM \frac{\sigma_{z}}2,
 \label{HSurfaceM}
\ee
where $J$ is the exchange coupling strength. Since $\cH$ is a sum of anti-commuting matrices we can write the spectrum down immediately
\be
E_k = \pm \sqrt{v^{2}k^{2} + (JM/2)^{2} },
 \label{ESurfaceM}
\ee
where $k^{2}=k_{x}^{2}+k_{y}^{2}$. 

We see that a gap of size $JM$ has opened up. However, even though from a theoretical standpoint this proposition seems promising, experimentally it has proven very difficult to fabricate a sample with
the requisite properties. Two key challenges need to be overcome: first, to observe most of the interesting surface phenomena, one requires the surface to remain insulating; however most ferromagnets in nature are metallic. Second, for a ferromagnet in a thin-film geometry, the magnetization vector usually lies in the plane, whereas  a perpendicular magnetization is required to open up a gap in the TI surface state.
To the best of our knowledge, this has yet to be achieved in an experiment, although some theoretical work has been done on this topic.\cite{Liu}

\subsection{Bulk Doping}

If surface doping with magnetic impurities fails, it is natural to try bulk doping. In ARPES experiments\cite{ChenMassiveDirac,WrayCoulomb} it was found that doping the bulk with non-magnetic impurities (such as Ca, Sn and Tl) did not result in a gap in the Dirac cone, as expected since they do not break TRI. Conversely, doping with magnetic impurities, for example \biseFe, resulted in a spectral gap that increased with the concentration of magnetic dopants $x$, with a gap of 60meV for $x=0.25$ (the bulk gap for \bise is $\sim0.3$eV). For the magnetic dopants Fe and Mn it was found that, at least for small $x$, the bulk was paramagnetic, while for the undoped samples the bulk was found to be diamagnetic. The magnetization measurements were not sensitive to the surface.

This raises the question of magnetic ordering in the bulk versus the surface. In general, ordered phenomena in lower dimensions are more fragile ($T_{c}^{3D} > T_{c}^{2D}$), for example, the XY model and Heisenberg models - in 1D they do not order at any temperature, in 2D they order only for $T=0$ and in 3D they order for $T<T_{c}$. This is also the case for superconducting order and general stability of lattices. However, in the case of a TI, we will argue that it is possible that $T_{c}^{\rm bulk}<T_{c}^{\rm surf}$. Therefore, there is a regime $T_{c}^{\rm bulk} < T < T_{c}^{\rm surf}$ in which the bulk is unordered (paramagnetic) and the surface is ordered (for example, ferromagnetic).

 To illustrate why this could be the case, first recall that magnetic ordering with the magnetization perpendicular to the TI surface implies opening of a gap in the spectrum of the surface states. This can be seen directly from Eqs.\ (\ref{HSurfaceM},\ref{ESurfaceM}). Now consider the ungapped surface spectrum, assuming half filling, so that the surface valence band is fully occupied and the conduction band is empty (Fig.~\ref{FigDiracCone}a). Gapping the surface states causes the occupied states to move down in energy (Fig.~\ref{FigDiracCone}b), so the total kinetic energy decreases. Therefore, the formation of a surface gap is favourable.
If the chemical potential is shifted either up or down then the net gain in kinetic energy is diminished and we expect $T_{c}^{\rm surf}$ to decrease.

Contrast this with the situation in the insulating bulk, which is gapped to begin with. In an ordinary insulator with negligible spin-orbit coupling it is not possible to generate magnetization in the initially spin-degenerate bands
without first closing the gap. Equivalently, one may recall that the spin susceptibility of an ordinary insulator with a negligible spin-orbit coupling vanishes. In the bulk of a topological insulator the situation is more complicated as a result of the strong spin-orbit coupling that is necessary for the occurrence of the topological phase. In this case, magnetic susceptibility can be significant,\cite{YuQAH} and can lead to bulk magnetic states at non-zero temperatures. Nevertheless, in this study we find that generically the surface critical temperature for the formation of magnetic order exceeds the bulk critical temperature.

Equivalently, we can imagine integrating out the electrons, and consider that the range of the RKKY interaction between the dopants is inversely proportional to the gap. Since the gap is small for the surface and large for the bulk, the range of the interaction is much longer for the surface, and hence long range order is expected to be stronger on the surface.

\section{Formalism}
\subsection{Mean field theory for the bulk system}

We consider first just the 3D bulk of the material, and study the coupling of electrons to magnetic dopants within a mean field approximation. We assume that the density of impurities is low enough that we can neglect impurity-impurity interactions and that there is no clustering. This is supported by experiments\cite{HorDopedBi2Te3} on doped \bite. Therefore, we add an on-site electron-impurity exchange interaction term\cite{YuQAH}
\be
 H =H_{e} + J \sum_{I} \vex{S}_{I} \cdot \vex{s}_{I}
\ee
where $H_{e}$ is the Hamiltonian for the electrons in a TI to be discussed in Sec. IV below. $J$ is an exchange coupling constant, the sum extends over all impurity sites $I$, $\vex{S}_{I}$ and ${\vex{s}_{I}}$ are the spin operators of the impurity and electron on site $I$, respectively. 

We define the average magnetization of the impurities and electrons as $\vex{M}$ and $\vex{m}$. The impurity magnetization is averaged over all impurities and over all sites. This virtual crystal approximation can be pictured as ``spreading out'' the localized magnetic moment of the impurities, so that there is a moment on all sites. We assume that the fluctuations around the mean are small, and can be neglected, so that the interaction term is decoupled
\be
\label{dec}
J \sum_{I} \vex{S}_{I} \cdot \vex{s}_{I}
  \simeq  J \vex{M} \cdot \sum_{i} \vex{s_{i}} + J \vex{m} \cdot \sum_{I} \vex{S_}{I} - NJ \vex{M} \cdot \vex{m}
\ee
where $N$ is the number of sites. We assume that the magnetization is in the $z$ direction, so $\vex{M} = M \hat{z}$ and $\vex{m} = m \hat{z}$. Then the mean-field Hamiltonian becomes,  $H^{MF} =  H_{e}^{MF} + H_{I}^{MF} -N J M m$, where
\be
 H_{e}^{MF} &=& \sum_{\vex{k}} \Psi_{k}^{\dagger}  \left[ \cH_{e}(\vex{k}) + J M \frac{\sigma_{z}}2- \mu  \right] \Psi_{k} \\
H_{I}^{MF} &=& J m \sum_{I} \vex{S}^{z}_{I}, \nonumber
\ee
and $\cH_{e}(\vex{k})$ is a matrix in spin and orbital spaces.
Since the mean field Hamiltonian is decoupled, we can write the energies as a sum
\be
E(\vex{k}, \lambda_{I}) = E_{e}(\vex{k}) + E_{I} - NJMm.
\ee
The first term reflects the energies of the mean field electron Hamiltonian $H^{MF}_{e}$. The second term is a sum over the expectation values of the spin operator in the $z$ direction at each impurity site $E_{I} = J m \sum_{I} \lambda_{I}$,  where $\lambda_{I}$ is the component of the spin in the $z$ direction on impurity $I$. The last term gives the overall shift in energy  following from the mean-field decoupling Eq.\ (\ref{dec}). 

To find the magnetizations, we calculate the expectation values of the electron and impurity spins in the ensemble defined by $H^{MF}$, and find the equations
\be
\label{M3DMagInfinite}
M 
 &=& -2Sx B_{S}(\beta J m S), \\
m &=& \frac1{N}  \sum_{\vex{k},i} \left(U^{\dagger}  \frac{\sigma_{z}}2 U \right)_{ii} f(E_{i}), \nonumber
\ee
where $B_{S}(y)$ is the Brillouin function\cite{AshcroftMermin} 
\be
B_{S}(y)
=
\frac{2S+1}{2S} \coth \frac{2S+1}{2S}y - \frac1{2S}\coth \frac1{2S}y.
\ee
$U$ is the matrix that diagonalizes $H^{MF}_{e}$ (it is a function of $J$, $M$, and $\vex{k}$) and $f(E)$ is the Fermi function. The chemical potential $\mu$ can be determined by summing over the average occupation for each energy (given by the Fermi-Dirac distribution) and equating to the total number of states. The number of states per site is
$n = \sum_{k} f(E_{e})/N$. For example, for half filling ($\mu=0$) at $T=0$ for a model with four states per site, we have $n=2$ since only half of those are occupied. The equations for $m$, $M$ and $\mu$ are coupled non-linear equations which can be solved self consistently by an iterative procedure.

\subsection{Adding the surfaces}

\begin{figure}[t]
\begin{center}
\includegraphics[width=2.5in]{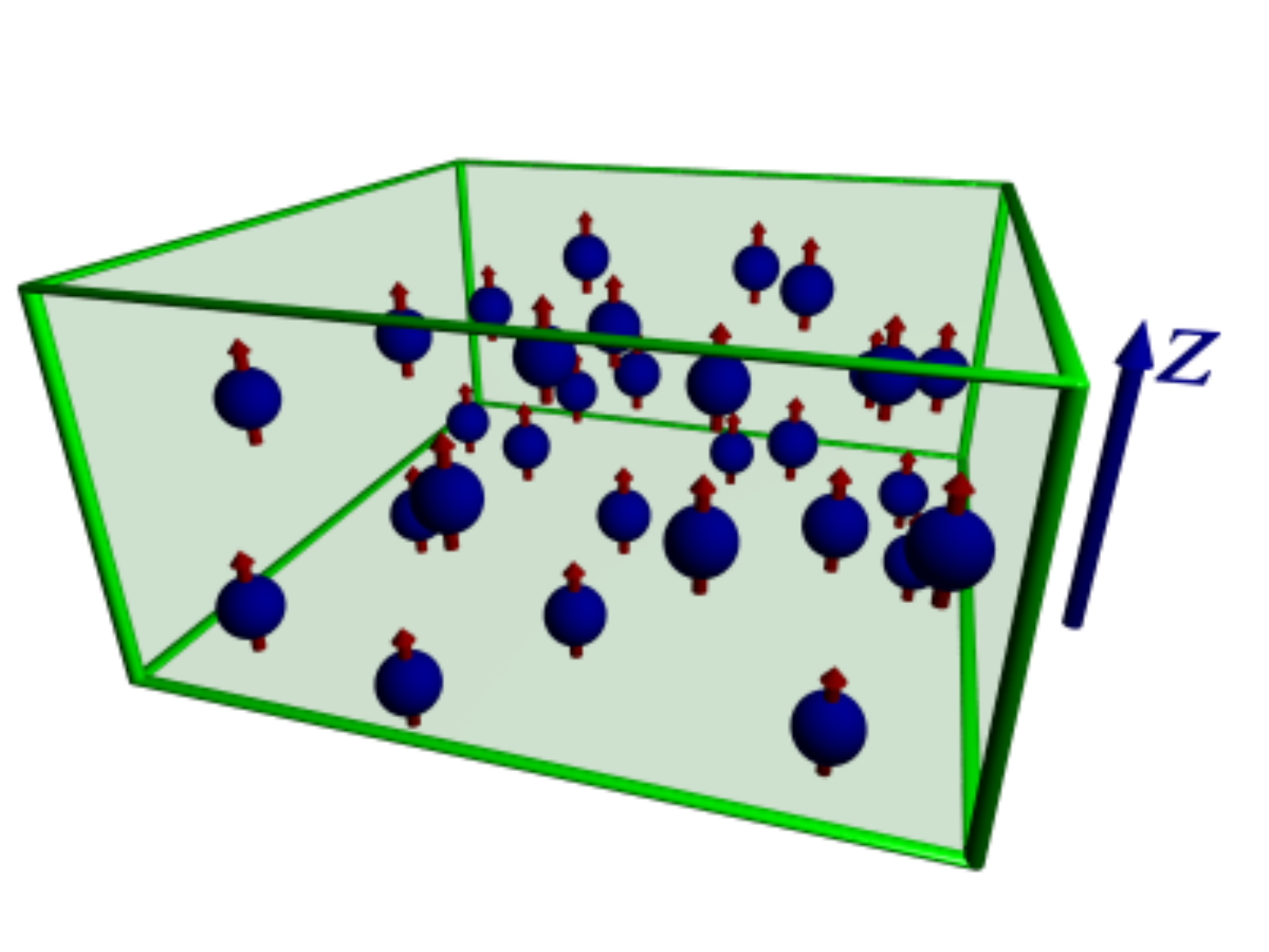}
\caption{A 3D TI in a slab geometry with a low density of randomly positioned magnetic impurities.}
\label{FigTISlab}
\end{center}
\end{figure}

Consider now a system with open boundary conditions in the $z$ direction and periodic boundary conditions in the $x$ and $y$ directions (see Fig.~\ref{FigTISlab}). The advantage of this setup is that in addition to the bulk, we can also investigate the magnetization near the two surfaces of the sample. We now take the real space Hamiltonian, and Fourier transform it only in $x$ and $y$, keeping the dependence on $z$ in real space. The problem is now a 1D problem in $z$, as opposed to a 0D problem before.

Once again, we rewrite the interaction term and then define the magnetizations as $z$-dependent fields. The mean-field Hamiltonian is now $H^{MF} = H_{e}^{MF} + H_{I}^{MF} - N_{\perp}J \sum_{j} M_{j} \cdot  m_{j}$, where $j$ labels the layers and 
\be
 H_{e}^{MF}  &=&  \sum_{\vex{k_{\perp}}} \Psi^{\dagger} _{k_{\perp}} \left[ \cH_{e}(\vex{k_{\perp}}) + J \vex{M} \otimes \frac{\sigma_{z}}2 - \mu \right]   \Psi_{k_{\perp}}, \nonumber \\
H_{I}^{MF} &=& J \sum_{j,I \in j} m_{j} \cdot \vex{S}_{I}.
\ee
Here $\cH_{e}(\vex{k_{\perp}})$ is a matrix in indices labeling the $z$ coordinate in addition to spin and orbital spaces.
$\vex{M}$ is a vector of length $L_{z}$ (the number of sites in the $z$ direction), $N_{\perp}=N/L_{z}$ is the number of sites in the $x$-$y$ plane, and by $\vex{M} \otimes \sigma_{z}$ we mean that $\vex{M}$ is expanded to a diagonal matrix of dimension $L_{z}$ and then we perform an external product with $\sigma_{z}$. We now diagonalize the full electron Hamiltonian $\cH_{e}(\vex{k_{\perp}}) + J \vex{M} \otimes \sigma_{z}/2 - \mu $, and find a matrix $U$ which diagonalizes it. 

We calculate the electron and impurity magnetizations of each layer, which are given by the sum over the expectation values of the spin operators
\be
\label{M3DMagInfinite}
	M_{i}  &=& -2Sx B_{S}(\beta J m_{i} S), \\
	m_{i} &=& \frac1{N_{\perp}}\sum_{l,j} \left(U_{il}^{\dagger}\frac{\sigma_{z}}2U_{il} \right)_{jj} f(E_{j}), \nonumber
\ee
where $U_{il}$ is a $d\times d$ block of $U$, and $d$ is the dimension of the Hamiltonian in the bulk ($d=4$ for \bise and $d=8$ for perovskite, the models are defined below). The chemical potential can be determined as before, by solving  $n = \sum_{\vex{k_{\perp}},\alpha} f(E_{\alpha}(\vex{k_{\perp}})) / N$ and $\alpha$ runs from 1 to $L_{z}d$. 

\subsection{Estimate of the surface critical temperature}

Consider a ferromagnetically ordered 2D surface of a TI. We can make a rough estimate of the critical temperature. The effective Hamiltonian is given by (\ref{HSurface}) and (\ref{HSurfaceM}), and the energies are given by (\ref{ESurfaceM}). The impurity and electron magnetizations are described by the coupled equations
\be
	M &=& -2Sx B_{S}(\beta J m S) \\
	m &=& \frac1{NJ} \sum_{k} \frac{\partial E_k}{\partial M} f(E_k-\mu) \nonumber
\ee
For $T \to T_{c}^{\rm surf}$ we have $m,M \to 0$, so we expand the above equations to first order in $M$ and $m$. We make use of $\partial E / \partial M = J^2 M / 4 E$ and the assymptotic form of the Brillouin function $B_{S}(y) \simeq y(S+1)/3S$ for $y \ll 1$, and find
\be
	M &=& -\frac23 S (S+1) \beta J x m,  \\
	m &=& -\frac{JM}{4N} \sum_{k} \frac1{E_k} [ 1 - f(E_k+\mu) - f(E_k-\mu) ]. \nonumber
\ee
We can evaluate the sum by converting it into an integral and imposing a momentum cutoff $\Lambda$. Assuming $E_k \simeq \pm v|k|$ (since $M \to 0$ at the critical point) we find a simple result\cite{Liu,Jungwirth} for exact half filling, i.e\ $\mu=0$, 
\be\label{tcsurf}
  T_{c}^{\rm surf}
 \simeq
\pi \frac{S(S+1)}{3 k_{B}} 
   \left( \frac{\Lambda a}{2\pi} \right)^{2} \frac{J^{2} x}{(\Lambda v)}.
\label{TcSurfEstimate1}
\ee
For $S=5/2$, $J=0.5$eV, $x=0.05$, $v=2\lambda_{\perp}a$ and a cutoff $\Lambda a / (2\pi) \simeq 1/5$ we find $T_{c}^{\rm surf} \simeq 100$K.
Away from half filling, when $\beta\mu \ll 1$, there is a small correction $\sim -\mu^2/(4k_Bv\Lambda)$ to the result in Eq.\ (\ref{tcsurf}).

A simple result can also be found for the case $\beta\mu \gg 1$ which is relevant for most values of $\mu$ inside the bulk gap,
\be
  T_{c}^{\rm surf}
 \simeq
\pi \frac{S(S+1)}{3 k_{B}} 
   \left( \frac{\Lambda a}{2\pi} \right)^{2} \frac{J^{2} x}{(\Lambda v)^{2}} (\Lambda v - |\mu|).
\label{TcSurfEstimate2}
\ee
As expected, $T_{c}^{\rm surf}$ is seen to decrease away from half filling.
\begin{figure}
\centering
\hspace{-0.4in}
\includegraphics[width=2.6in]{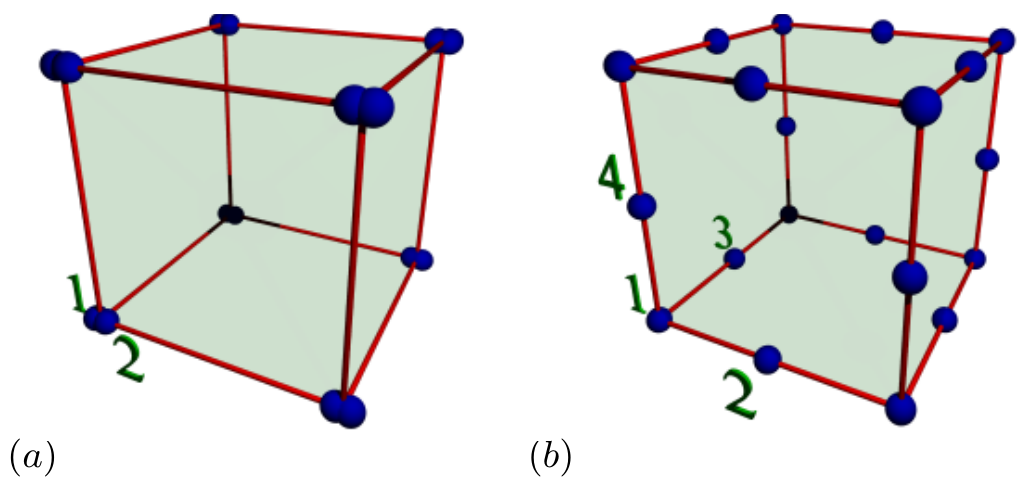}
\caption{(a) A unit cell of the discretized \bise model, a simple cubic lattice with two orbitals per site.
(b) A unit cell of the perovskite lattice, which can be described as edge centered cubic. The four basis sites are labelled. }
\label{FigLattices}
\end{figure}

To complete the argument we would now like to give a similar simple estimate for  $T_{c}^{\rm bulk}$. Unfortunately, the bulk critical temperature is not easy to estimate, since unlike the topologically protected surface state, whose physics is simple and universal, the bulk of a TI can be complicated and $T_{c}^{\rm bulk}$ will generally depend on the details of the band structure and other factors. For  Bi$_2$Se$_3$  with 5\% concentration of Cr dopants, Ref.\ \onlinecite{YuQAH} estimates $T_{c}^{\rm bulk}\simeq 70$K using first-principles numerical calculations. Comparing with our rough estimate for $T_{c}^{\rm surf}$ given above we see a clear indication that a $T_{c}^{\rm bulk}<T_{c}^{\rm surf}$ regime can be easily obtained.
\begin{figure*}
\centering
\includegraphics[width=7in]{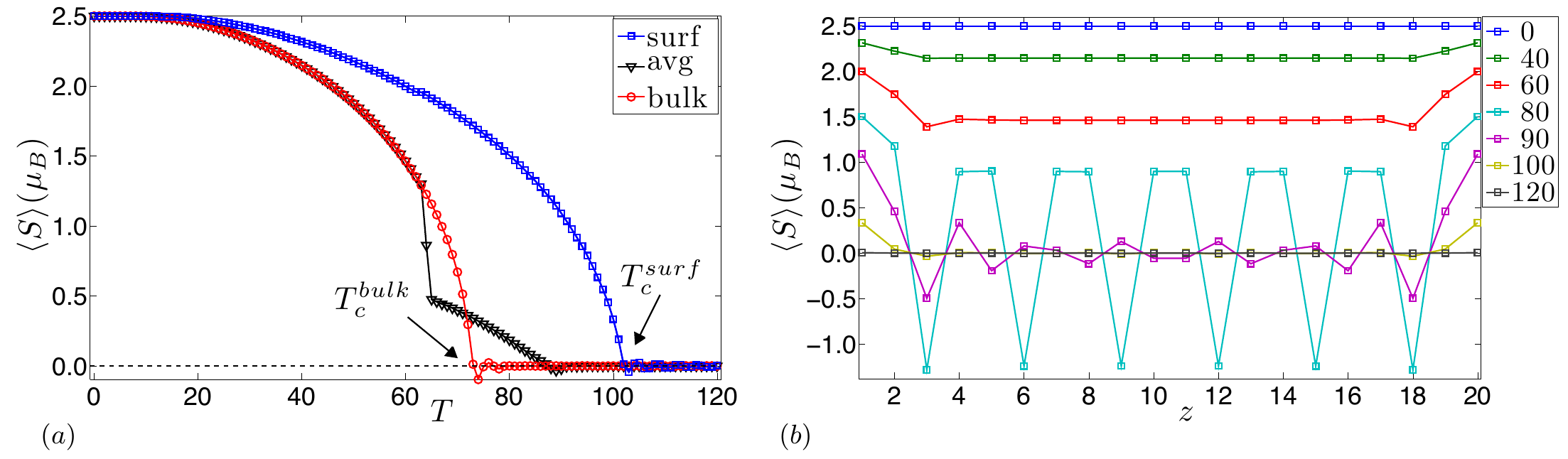}
\caption{(a) Impurity spin expectation value as a function of temperature for the discretized \bise lattice. We plot the order parameter on the surfaces (``surf''), the averaged bulk result for the 14 middle sites (``avg''), and the results of the separate bulk calculation (``bulk''). (b) Impurity spin expectation value as a function of the $z$ coordinate, for different temperatures. For both plots the parameters are $L_{x}=L_{y}=40$, $L_{z}=20$, $S=5/2$, $J=0.5$eV, $x=0.05$, and the rest of the parameters are chosen to fit the \bise dispersion close to the $\Gamma$ point based on first principles calculations:\cite{ZhangSingleDiracCone, LiDynamicalAxion, FuOddParityTSC}  $\gamma_{0}=0.3391\eV,
\gamma_{\perp}=0.0506\eV, \gamma_{z}=0.0717\eV, \epsilon = 1.6912\eV, t_{\perp}=0.3892\eV, t_{z}=0.2072\eV, \lambda_{\perp}=0.2170\eV$ and $\lambda_{z}=0.1240\eV$.  }
\label{FigPlotMBi2Se3}
\end{figure*}

\section{Models and Results}

Below we present our results for the two tight binding models we considered in this study: a model for \bise regularized on a simple cubic lattice\cite{FuOddParityTSC} and a model on the perovskite lattice\cite{Weeks}. These simple tight-binding models with spin-orbit coupling exhibit non-trivial topological invariants for a broad range of model parameters and have been used widely in the literature to study the physical properties of topological insulators.

\subsection{Effective model for  \bise regularized on the cubic lattice}

We consider electrons hopping on a simple cubic lattice with two orbitals per site (Fig.~\ref{FigLattices}a). The form of the Hamiltonian and the parameters (see caption Fig.\ref{FigPlotMBi2Se3}) were chosen to fit the dispersion near the center of the Brillouin zone obtained by the first principles calculation \cite{ZhangSingleDiracCone, LiDynamicalAxion, FuOddParityTSC} for \bise. The electron Hamiltonian is given by
\be
\cH_{e}(\vex{k}) &=& d_4 + \sum_{\mu} d_{\mu} \Gamma_{\mu} \\
d_0 &\equiv& m_k = \epsilon - 2 \sum_i t_{i} \cos(k_i a_{i}) \nonumber \\
d_i &\equiv& -2 \lambda_{i} \sin(k_i a_{i}) \nonumber \\
d_{4} &\equiv& \gamma_{0} - 2\sum_{i}\gamma_{i} \cos(k_{i}a_{i}) \nonumber
\ee
where we have chosen the gamma matrices so that $\Gamma_0 = \tau_{1}\otimes\sigma_0$, $\Gamma_1 = -\tau_3 \otimes \sigma_2$, $\Gamma_2 = \tau_3 \otimes \sigma_1$, $\Gamma_3 = \tau_2 \otimes \sigma_0$, and $\tau_{i}$ are Pauli matrices in the orbital subspace and $\sigma_{i}$ are Pauli matrices in the spin subspace. The energies are
\be
E_{e}
&=& d_{4} + \gamma \sqrt{d^{2}_{1} + d^{2}_{2} + \left[ \sqrt{d^{2}_{0}+d_{3}^{2}} + \delta \tilde{M} \right]^{2} }
\ee
where $\tilde{M} \equiv JM/2$, and $\gamma,\delta=\pm1$. Note that if there are no impurities ($J=0$) then we get the ``clean'' doubly degenerate electron spectrum $E_{0} = d_{4} \pm \sqrt {\sum_{\mu} d_{\mu}^{2}}$.

The results for the magnetization in this model are presented in Fig~\ref{FigPlotMBi2Se3}. The bulk is ferromagnetically ordered up to $T_{c}^{\rm bulk}\simeq 73$K, and the surface remains FM ordered up to $T_{c}^{\rm surf}\simeq 102$K for the two surfaces (see Fig~\ref{FigPlotMBi2Se3}a). Therefore, the window in which the surface is ordered and the bulk is unordered (paramagnetic) is $\simeq 29$K.
We plot the magnetization in the $z$ direction in Fig~\ref{FigPlotMBi2Se3}b. Here the effect can be seen clearly. If we ramp up the temperature from $T=0$ at first all spins, regardless of if they are in the bulk or surface, are fully polarized. As we increase the temperature, the magnetization of the bulk drops faster than the magnetization of the surface. Eventually we cross $T_{c}^{\rm bulk}$ at which stage the magnetization of the bulk is zero, but the magnetization of the surface is finite. If we increase the temperature further, the magnetization of the surface drops, but the magnetization of the bulk remains at zero. Once we cross $T_{c}^{\rm surf}$, the magnetization of both the bulk and the surface vanish -- thermal fluctuations have broken the ordered phases.  

In samples with open surfaces we observed spatial fluctuations in the bulk magnetization as the temperature approached $T_{c}^{\rm bulk}$ from above, e.g. the $T=80$K curve in  Fig~\ref{FigPlotMBi2Se3}b. We attribute these to the large magnetic susceptibility of the bulk, which diverges at $T\to T_{c}^{\rm bulk}$, and the proximity of the ordered surfaces.
The correct $T_{c}^{\rm bulk}$ can be obtained from the bulk calculation with periodic boundary conditions, the results of which are plotted in Fig.~\ref{FigPlotMBi2Se3}a.

\begin{figure}
\includegraphics[width=3.3in]{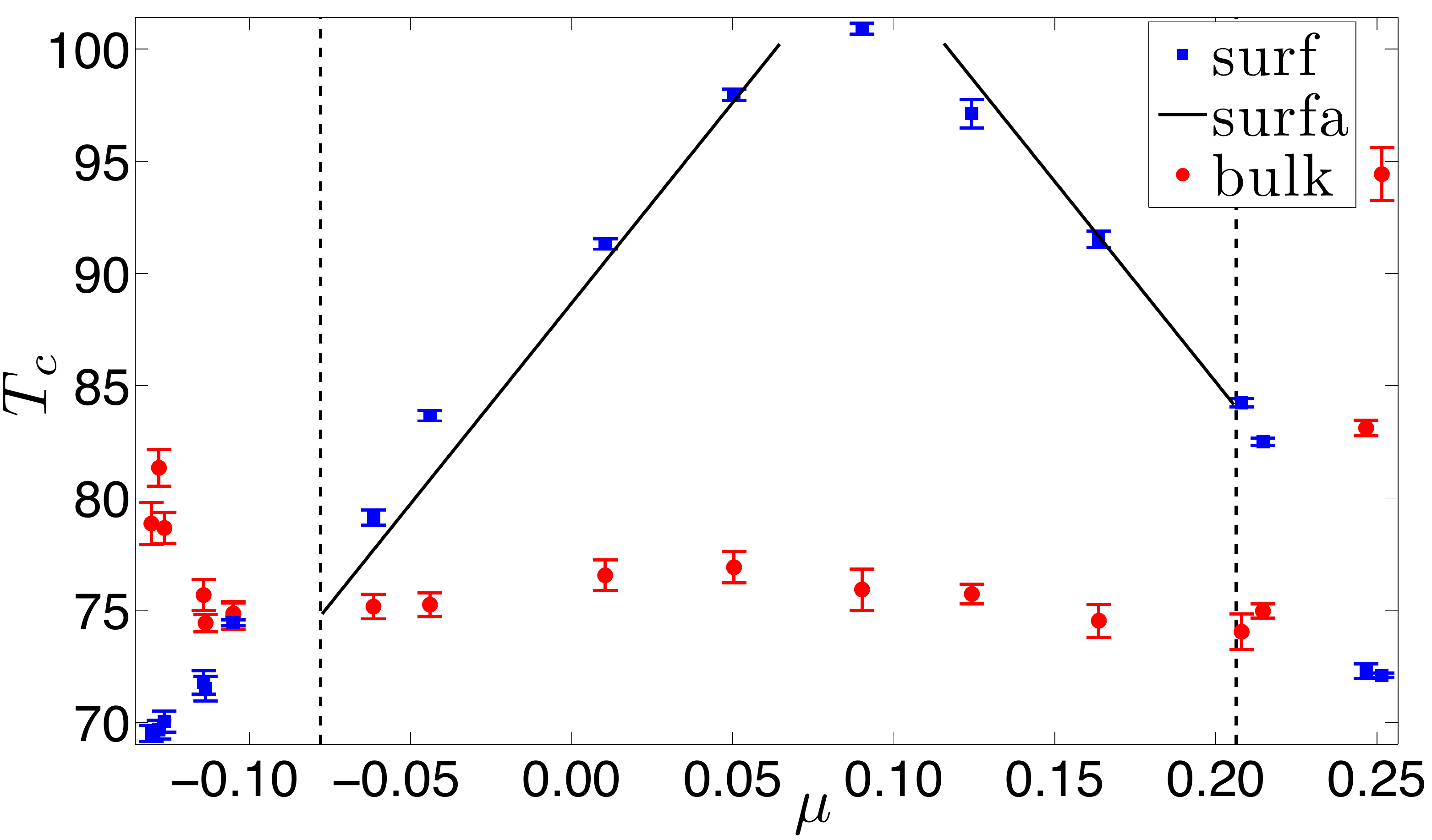}
\caption{Critical temperature as a function of the chemical potential $\mu$ for the discretized \bise lattice. We plot the surface critical temperature (squares, blue), and the approximate surface temperature from Eq.~(\ref{TcSurfEstimate2}) (solid line, black). In addition, we plot the bulk critical temperature as obtained from an average over the bulk sites (circles, red). The error bars were obtained by fitting the magnetization below the critical temperature to the known behaviour for a second order phase transition, $M \propto (T_{c}-T)^{1/2}$. The parameters are as in Fig.~\ref{FigPlotMBi2Se3}, except for $L_{z} = 10$. The cutoff was chosen for best fitting by eye $\Lambda a / 2 \pi = 0.215$. The vertical dashed lines mark the bulk gap.  }
\label{FigPlotTcmuBi2Se3}
\end{figure}

We plot the critical temperature for the surface and the bulk as a function of the chemical potential in Fig~\ref{FigPlotTcmuBi2Se3}. The surface critical temperature is maximal when the chemical potential intersects the surface Dirac point, which happens for $\mu\simeq 0.09$eV, and falls on both sides, as expected. The result agrees well with the predicted linear dependence  of the surface critical temperature on $\mu$ predicted in Eq.\ (\ref{TcSurfEstimate2}). The bulk critical temperature is approximately constant within the bulk gap, as expected -- since there are no bulk states within the gap, changing the chemical potential should not change the critical temperature. 

\subsection{Perovskite lattice model}

To ascertain whether the results obtained for the \bise model are in fact generic, we investigate another simple lattice  model known to give a robust TI.
The perovskite lattice, or edge-centered cubic lattice, consists of atoms on a simple cubic lattice with a four point basis, corresponding to atoms at the center of each edge (Fig.~\ref{FigLattices}b). Tight binding electrons on this lattice (with spin orbit coupling) were recently shown to form a TI.\cite{Weeks} The momentum-space electron Hamiltonian, written in the basis of four sublattice sites indicated in Fig.~\ref{FigLattices}b, is $\cH_{e} = \cH_{0} + \cH_{SO}$. Here $\cH_{0}$ is the hopping part
\be
\cH_{0} = -2t
\begin{pmatrix} 
		 0  & \cos k_{x} & \cos k_{y} & \cos k_{z}  \\ 
		\cos k_{x} & 0 & 0 & 0 \\
		\cos k_{y}  & 0 & 0 & 0 \\
		\cos k_{z} & 0 & 0 & 0
\end{pmatrix}
\ee
and $\cH_{SO}$ is the spin orbit part, $\cH_{SO} = \sum_{\mu} \sigma_{\mu} \otimes \cH_{SO}^{\mu}$, and 
\be
\cH_{SO}^{x} = 4 \lambda i \sin k_{y}  \sin k_{z} 
\begin{pmatrix} 
		 0 & 0 & 0 & 0  \\ 
		 0 & 0 & 0 & 0 \\
	 	 0 & 0 & 0 & 1 \\
		 0 & 0 & -1 & 0
\end{pmatrix},
\ee
and similarly for $y$ and $z$. 

\begin{figure*}
\centering
\includegraphics[width=7in]{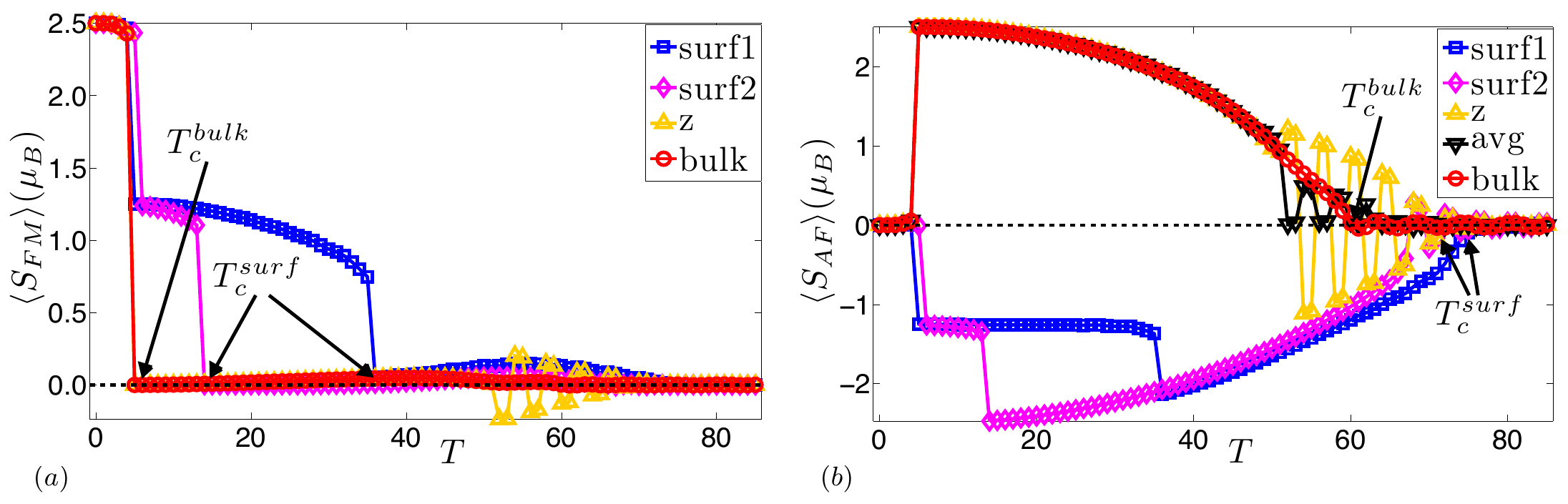}
\caption{(a) Ferromagnetic  order parameter as a function of temperature for the perovskite lattice, with parameters $L_{x}=L_{y}=20$, $L_{z}=10$, $S=5/2$, $J=0.25\eV$, $t=1\eV$, $\lambda=0.2\eV$, $x=0.05$. We plot the order parameter on each of the surfaces (``surf1'' and ``surf2''), the result for the site at $z=L_{z}/2$ (``z''), and the results of the separate bulk calculation (``bulk''). (b) Antiferromagnetic  order parameter as a function of temperature for the perovskite lattice, with parameters as in (a). Here we also plot the average over the four central sites (``avg''). }
\label{FigPlotMPerovskite}
\end{figure*}

The results for magnetization in this model are presented in Fig~\ref{FigPlotMPerovskite}. In this case we observe that for a large range of temperatures, the magnetization on basis sites 1 and 4 is similar and opposite in sign from the magnetization on basis sites 2 and 3 (which are equal due to a rotational symmetry around the $z$ axis). This motivates the definition of a FM order parameter $S_{i}^{\rm FM}=(S_{i,1}+S_{i,2}+S_{i,3}+S_{i,4})/4$ and an antiferromagnetic (AF) order parameter $S_{i}^{\rm AF}=(S_{i,1}-S_{i,2}-S_{i,3}+S_{i,4})/4$ (where $S_{i,\ell}$ is the expectation value of the impurity spin on basis site $\ell$ of lattice site $i$). We see that the system is ferromagnetically ordered up to $T\simeq 4.5$K, where it becomes AF ordered. The bulk remains AF ordered until $T_{c}^{\rm bulk}\simeq 60.5$K, and the surface remains AF ordered until $T_{c}^{\rm surf}\simeq 72$K and 78K for the two surfaces. Therefore, the maximum window in which the surface is ordered and the bulk is unordered (paramagnetic) is $\simeq 17.5$K. Note that the difference in surfaces results from the fact that the unit cell is not symmetric under reflection along $z$ - the ``top'' of the system ends with basis site 4 and the bottom ends with basis site 1 (see Fig.~\ref{FigLattices}a). In addition, we see that the surfaces undergo an additional partial phase change signified by discrete jumps of the order parameter seen at $T \simeq 4.5,5.5,13.5,35.5$K, which we claim as additional evidence that the bulk and surface differ. 

\section{Conclusions}

We have demonstrated that magnetically doped topological insulators can have a sizeable window where the bulk is paramagnetic and the surface is magnetically ordered. Our conclusions are based on general arguments that involve only universal properties of the topologically protected surface states and on numerical calculations performed on simple lattice models of 3D topological insulators. Physically, these results are in accord with the intuition that the metallic state on the surface should be more susceptible to magnetic ordering than the insulating bulk, although this expectation is only partially borne out in systems with strong spin-orbit coupling. 

The results reported in this study rely on two key approximations: (i) the mean-field decoupling  of the exchange interaction between electrons and magnetic dopants indicated in Eq.\ (\ref{dec}), and (ii) the `virtual crystal' approximation which replaces the localized magnetic moments of dopant atoms by their average over all lattice sites. We have attempted to bypass the latter approximation by solving the problem in real space without averaging over sites. This procedure gives reasonable results for $T_{c}^{\rm bulk}$ compared to those obtained in the virtual crystal but is numerically more costly. The system sizes that we could simulate did not allow us to unambiguously determine the surface critical temperature. Nevertheless based on these results we feel that, in a large crystal, (ii) will not lead to a significant error in the determination of the critical temperatures. 

Since fluctuations around the mean field result are typically stronger in 2D than in 3D, they will likely reduce the size of the temperature window between  $T_{c}^{\rm bulk}$ and $T_{c}^{\rm surf}$. 
On the other hand, our mean field calculation did not include electron-electron interactions, which would tend to stabilize the long range magnetic order and thereby strengthen the ordered phases. It has been shown recently that electron-electron interactions can lead to spontaneous breaking of TRI on the surface of a TI,\cite{BaumMagneticInstability} even in the absence of magnetic dopants. Thus the interactions might in fact strengthen the effect found in our study, although a more detailed investigation would be needed in order to obtain a quantitative result. Overall, in our opinion it is very likely that the combined effect of the magnetic dopants and the electron-electron interactions can account for the experimentally observed surface excitation gap without bulk magnetic order in Mn and Fe doped \bise. 

We note that the critical temperature for the bulk magnetic ordering of the magnetically doped topological insulator Bi$_{2-x}$Mn$_x$Te$_3$ was recently measured\cite{HorDopedBi2Te3} to be $9-12$K for $x=0.04-0.09$. This critical temperature is smaller than our results and those of  Ref.\ \onlinecite{YuQAH}, which estimates $T_{c}^{\rm bulk}\simeq 70$K for Cr-doped \bise using first-principles numerical calculations. We chose our coupling constant $J$ based on the first-principles results, so we expect that for a reduced coupling constant the temperature window would shift to lower temperatures. As noted above, our arguments are universal, hence the exact details of the material and coupling might alter the result quantitatively but should not change it qualitatively. Taking the above finding for Bi$_{2-x}$Mn$_x$Te$_3$ as a guideline, one may surmise that $T_{c}^{\rm bulk}$ for Fe and Mn doped \bise lies in a similar range of temperatures. It is then entirely possible that the ARPES experiments,\cite{ChenMassiveDirac,WrayCoulomb} performed at $\sim 20$K, detected surface magnetic ordering without bulk magnetism as advocated in this paper. Careful surface-sensitive magnetic measurements, as proposed recently,\cite{vazifeh1} might be able to probe this intriguing phenomenon directly.


\begin{acknowledgements}
The authors have benefited from discussions with L. Fu, L. Levitov and C. Weeks. This work was supported by NSERC and CIfAR.
\end{acknowledgements}


\end{document}